\documentclass{pasj00}
\draft

\begin{document}
\SetRunningHead{R. Fukuoka \etal}{Discovery of G\,0.174$-$0.233}
\Received{2009/02/11}
\Accepted{2009/03/06}

\title{Suzaku Observation Adjacent to \\the South End of the Radio Arc}
\author{Ryosuke~\textsc{Fukuoka}, 
Katsuji~\textsc{Koyama},\\ Syukyo G.~\textsc{Ryu}, and Takeshi Go~\textsc{Tsuru}}
\altaffiltext{}{Department of Physics, Graduate School of Science, 
Kyoto University, \\Kita-shirakawa Oiwake-cho, Sakyo, Kyoto 606-8502}
\email{fukuoka@cr.scphys.kyoto-u.ac.jp}

\maketitle

\begin{abstract}

Suzaku observed the Galactic center region near the Radio Arc at $\sim\timeform{20'}$ 
southeast of Sagittarius\,A$^*$.
In the \timeform{18'}$\times$\timeform{18'} field of view, 
we found four distinct X-ray sources: 
a bright star and a diffuse source associated with the star clusters in 
the soft band (0.5--2.0~keV), a small clump in a higher energy band (4--6~keV), 
and a peculiar clump in the 6.4~keV line band.
The latter two clumps are located at the south end of the Radio Arc. 
This paper reports on the results, and discusses the origin of these X-ray sources, 
with a particular emphasis on small clumps.

\end{abstract}

\section{Introduction}

Radio non-thermal filaments (NTFs; e.g. \cite{larosa00}) are unique structures 
seen only in the Galactic center (GC) region within a few degrees around Sagittarius\,A$^*$ (Sgr\,A$^*$).
Most of the NTFs are nearly perpendicular to the Galactic plane.
The most striking and large-scale NTF is 
the Radio Arc threading the Galactic plane at $l\sim\timeform{0.2D}$ (\cite{yusef84}).
The Radio Arc may be a site of high-energy activity:
acceleration to relativistic electrons in an enhanced magnetic field.

Another feature of the high-energy activity in the GC is 
diffuse X-ray emission in the 6.4~keV line from neutral irons (Fe\emissiontype{I}).
After the first discovery by ASCA in the Sgr\,B2 region \citep{koyama96}, 
several 6.4~keV clumps have been found in the regions of Sgr\,B1 \citep{nobukawa08}, Sgr\,C 
(\cite{murakami01}; \cite{nakajima09}), and near the Radio Arc \citep{yusef07}.
They have a large equivalent width (EW) of $\geq1$~keV for the 6.4~keV line, 
and positionally correlate with molecular clouds.
Some of them were also found to be time variable 
in flux and morphology (\cite{muno07}; \cite{koyama08}; \cite{inui09}).

\citet{koyama96} proposed that these 6.4~keV clumps are X-ray reflection nebulae (XRNe), 
which are molecular clouds irradiated by super-massive black-hole Sgr\,A$^*$ at the GC.
The required luminosity of Sgr\,A$^*$ to explain the flux of several XRNe 
is $\sim10^{39}$~ergs s$^{-1}$, 
but the present luminosity is only 2$\times10^{33}$~ergs s$^{-1}$ \citep{baganoff03}.
Thus Sgr\,A$^*$ had to be six orders of magnitude brighter 
about 300~years ago, the light traveling time from Sgr\,A$^*$ to the XRNe.
This scenario can also explain the time variability of some 6.4~keV clumps, 
assuming that the flare of Sgr\,A$^*$ was time variable. 

On the other hand, \citet{yusef07} proposed that the 6.4~keV emission 
comes from low-energy cosmic-ray electrons (LECRs) bombarding the clouds.
In this scenario, a possible correlation of NTFs and the 6.4~keV or
the other X-ray band emissions should be found, because NTFs may be sites 
of relativistic electrons, and hence LECRs may also be abundant.
However, no significant association between the Radio Arc
and the 6.4~keV line, or any other X-ray band flux has been found.

We therefore performed a Suzaku observation to search for the 
X-ray emission from the Radio Arc.
To observe any faint X-ray sources near the Radio Arc,
the bright Galactic center diffuse X-ray emission (GCDX; e.g. \cite{koyama07b})  
would be the most serious background.
Since GCDX decreases as the distance from Sgr\,A$^*$ increases (\cite{koyama89}, \cite{yamauchi90}), 
we selected the most distant region of the Radio Arc, 
adjacent to its south end.

This paper reports the observational results of the X-rays, particularly 
the 6.4~keV line emission adjacent to the south end of the Radio Arc.
Throughout this paper, we assume the distance toward the GC to be 8~kpc \citep{reid93}, 
and use the Galactic east and north as positive Galactic longitude and latitude, respectively.

\section{Observation}

We conducted a deep Suzaku observation centered at 
$(\alpha, \delta)_{\rm J2000.0}$=(\timeform{17h23m14s}, \timeform{-28D52'56''}) on 2007 August 31--September 3.
The position is in the southeast direction of the GC by $\sim\timeform{20'}$,
which is near the south-end of the Radio Arc.
The observation was performed with the X-ray Imaging Spectrometer (XIS: \cite{koyama07a}).

The XIS is equipped with four X-ray CCD camera systems, 
each placed on the focal planes of the four X-Ray Telescopes \citep{serlemitsos07}.
Three of the XIS sensors (XIS\,0, 2, and 3) use front-illuminated (FI) CCDs, 
and the remaining one (XIS\,1) has a back-illuminated (BI) CCD.
The FI CCDs have higher sensitivity and lower background than 
the BI in the energy band above 1--2~keV, 
while the BI CCD has higher sensitivity than the FI in the energy below $\sim1$~keV. 
After the unexpected anomaly in 2006 November, 
one of the FIs (XIS\,2) has not been functional.
The XIS field of view (FOV) covers on $\sim\timeform{18'}\times\timeform{18'}$ 
region with a point-spread function (PSF) 
of \timeform{1.9'}--\timeform{2.3'} in the half-power diameter 
and a pointing accuracy of up to $\sim\timeform{20''}$ \citep{uchiyama08}. 
The calibration sources of \atom{Fe}{}{55} are installed 
in the two corners of each CCD.

The observation was performed in the normal clocking mode 
with the spaced-row charge injection technique \citep{uchiyama09}.
Data were processed with version 2.1\footnote{See $<$http://www.astro.isas.jaxa.jp/suzaku/process/$>$.}
of the pipeline processing software package.  
We removed data during passages through the South Atlantic Anomaly,  
at Earth elevation angles below 4$^\circ$, 
and at Earth day-time elevation angles below 10$^\circ$.
We checked the raw count rate around these elevation angles, and verified
that the contaminating emission due to the solar X-ray
reflection is negligible.
After these filtering steps, the net exposure time was $\sim144$~ks. 
We analyzed the screened data using the HEADAS software version 6.5.1 
and XSPEC version 11.3.2\footnote{See $<$http://heasarc.gsfc.nasa.gov/docs/software/lheasoft$>$.} 
, and used the calibration databases released on 
2008 September 5\footnote{See $<$http://www.astro.isas.jaxa.jp/suzaku/caldb/$>$.}.
Using the calibration sources, 
the systematic gain uncertainty was found to be $\sim10$~eV at 5.895~keV.

\section{Analysis and Results}

\subsection{Images}

\begin{figure*}[!ht]
  \begin{center}
    \FigureFile(160mm,){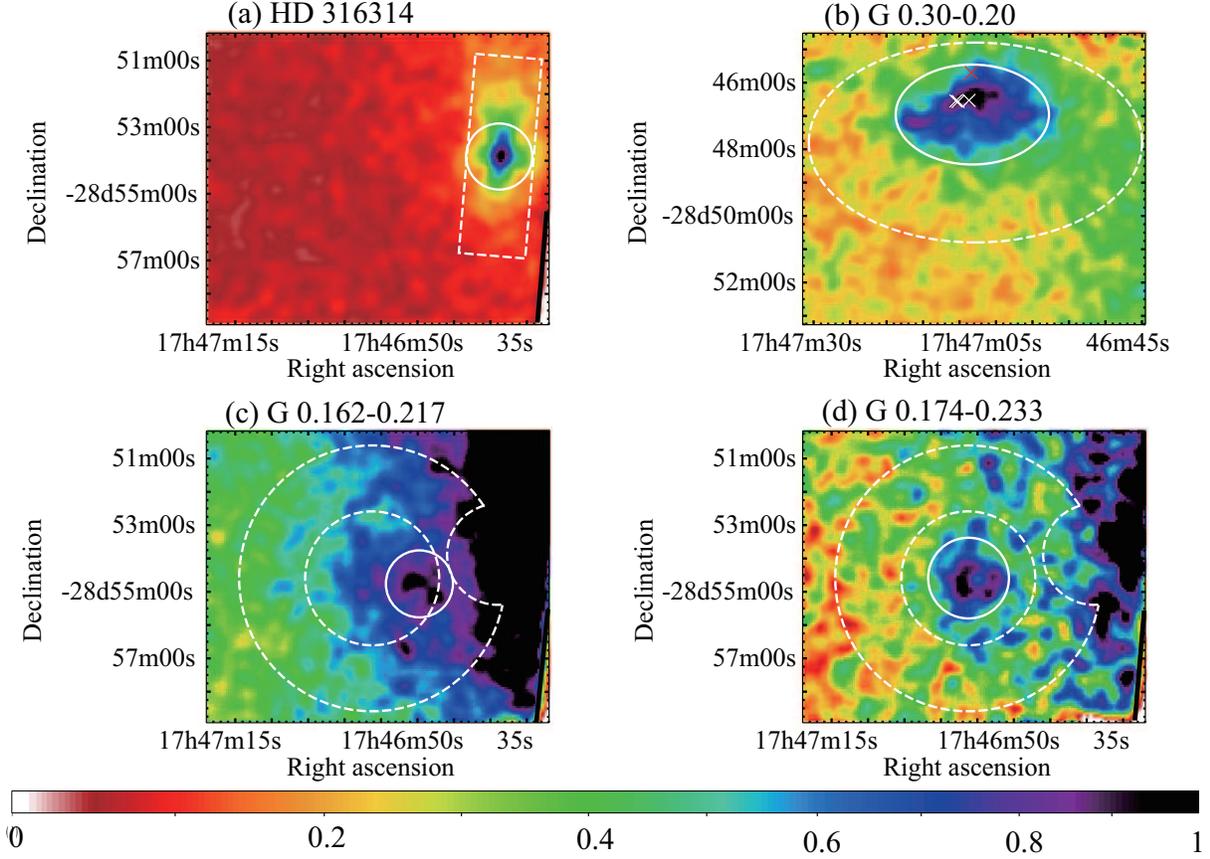}
\end{center}
  \caption{
		X-ray images of the Suzaku sources.			
		The fluxes of the three XIS data were co-added and 
		the color code is in log-scale with the normalized peak flux (peak= 1).
		The common color-map scale bar is also indicated.
       	The images were divided with the exposure maps 
		after subtracting the non--X-ray background and removing the calibration sources.
		The thick line is the edge of the XIS field of view.
		The solid and dashed line regions indicate the data-extraction area.
		(a) Soft X-ray band (0.5--2.0~keV) image of HD\,316314. 
		(b) Same as (a), but for G\,0.30$-$0.20. The white crosses are the X-ray stars inside
             the star clusters DB\,00-5/00-6, 
			while the red cross is the position of a Chandra transient source.
		(c) The 4.0--6.0~keV energy band image of G\,0.162$-$0.217 / Suzaku\,J174651$-$285435. 
			The strong emission near the west edge of the XIS field is the main part of the GCDX.  
		(d) The 6.4~keV-line band (6.3--6.5~keV) image of G\,0.174$-$0.233 / Suzaku\,J174656$-$28543 
			for the source and background, respectively.}
	\label{fig:image}
\end{figure*}

We made X-ray images in various energy bands,  
subtracted the non--X-ray background 
generated with the \texttt{xisnxbgen} package \citep{tawa08}, 
corrected for vignetting with the \texttt{xissim} package \citep{ishisaki07}, 
and smoothed the images with a Gaussian kernel with $\sigma$ $\sim\timeform{20"}$.
We then searched these multi-band images for X-ray sources. 
In the soft X-ray band (0.5--2.0~keV), we found an X-ray point source.
To determine the position, we made projection profiles of this source 
along the right ascension ($\alpha$) and declination ($\delta$) in the soft X-ray band.
With a Gaussian-plus-constant model fitting, we determined the peak position to be 
$(\alpha, \delta)_{\rm J2000.0}$ =(\timeform{266.6651D}$\pm \timeform{0.0003D}$, \timeform{-28.8968D}$\pm \timeform{0.0003D}$).
Within the Suzaku positional uncertainty of $\sim\timeform{20''}$, 
we found one Chandra source at $(\alpha, \delta)_{\rm J2000.0}$ = (\timeform{266.6629D}, \timeform{-28.8977D}), 
which has been catalogued as CXOGC\,J174639.0$-$285351, identified as an X-ray 
counterpart of HD\,316314 \citep{muno06}.
Since no other Chandra source was found in the error region, the Suzaku source is
most likely to be an X-ray counterpart of HD\,316314.
We then registered the Suzaku coordinate using this point source.
The Chandra frame is accurate to within \timeform{1''}\footnote{See $<$http://cxc.harvard.edu/proposer/POG/$>$.}, 
and hence we shifted the Suzaku coordinates by 
$\Delta(\alpha, \delta)$=(\timeform{0.0023D}, \timeform{-0.0009D}).
Hereafter, we use these fine-tuned coordinates for all Suzaku sources in this paper.

The Suzaku image of HD\,316314 \citep{muno06} in the 0.5--2.0~keV band is shown in figure~\ref{fig:image}a.
The Maltese-cross shape is due to the PSF of the XRT.
Consequently, this shape demonstrates that HD\,316314 is a point-like X-ray source (see also \cite{sugizaki09}).  
In the same X-ray band (0.5--2.0~keV) image, 
we also found a diffuse source, as is shown in figure~\ref{fig:image}b.
From the center position, we named this source G\,0.30$-$0.20.

In the higher energy band (4--6~keV), we can see an excess emission near the 
edge of the FOV at the northwest direction (figure~\ref{fig:image}c).
From the peak position, 
we refer to this source as G\,0.162$-$0.217/Suzaku\,J174651$-$285435 (see subsubsection 3.2.2).
The strong emission at the edge of the XIS field of view is due to 
the GCDX extending to the source from its flux peak at Sgr\,A$^*$.  

Interestingly, we found another clump in the
narrow energy band, the 6.4~keV-line (6.3--6.5~keV) band (figure~\ref{fig:image}d).
This clump is designated as G\,0.174$-$0.233/Suzaku\,J174656$-$285430 (see subsubsection 3.2.1).
Although the sizes of these two sources seem to be comparable to the Suzaku PSF, 
these are likely to be diffuse sources, because no hint of the Maltese-cross shape is found 
 (see subsection 3.2).

\subsection{Point-like or Diffuse}

The soft source HD\,316314 is surely point-like, while G\,0.30$-$0.20 is
diffuse or due to multiple sources.
The other two sources (G\,0.162$-$0.217/Suzaku\,J174651$-$285435 and G\,0.174$-$0.233/Suzaku\,J174656$-$285430) 
are controversial, and hence we examined whether these are point-like or diffuse.

\subsubsection{G\,0.174$-$0.233}

We first checked possible contamination from faint point sources 
while referring to the X-ray point-source catalog by \citet{muno06}, and 
found only one source (CXOGC\,J174651.9$-$285509) inside the source region.
This point source is $\sim\timeform{1'2}$ from the center of the G\,0.174$-$0.233
with the photon flux in the 0.5--8.0~keV energy band 
being 2.9$\times10^{-6}$~cm$^{-2}$~s$^{-1}$, 
which is only $\sim 6\%$ of that of G\,0.174$-$0.233.
We thus conclude that the contribution of the point sources is negligible.
We note that G\,0.174$-$0.233 was not detected with Chandra. 
Chandra (ACIS) has a lower efficiency and higher background than Suzaku (XIS) 
to detetct a faint diffuse source, such as G0.174$-$0.233.
The exposure time of Chandra ($\sim10$~ks) was also much shorter than that of Suzaku ($\sim140$~ks).

\begin{figure}[!ht]
  \begin{center}
    \FigureFile(80mm,80mm){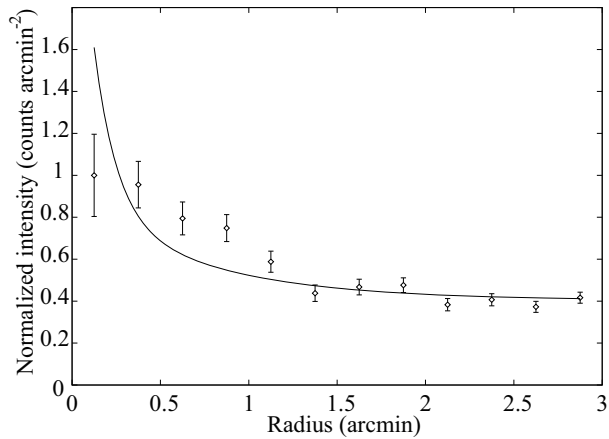}
  \end{center}
  \caption{Radial profile of G\,0.174$-$0.233 around its peak position in the 6.3--6.5~keV energy band.
	The observed fluxes and 1$\sigma$ errors are shown with diamonds and bars, respectively.
	The flux is normalized to its maximum value.
	The solid line shows the best-fit results with the PSF-plus-constant model.
  }\label{fig:radpro-64cl}
\end{figure}

We then compared the radial profile of G\,0.174$-$0.233 around the peak position with a PSF.
The peak position of G\,0.174$-$0.233 was determined in the same way as HD\,316314, 
and was found to be $(\alpha, \delta)_{\rm J2000.0}$=(\timeform{266.73665D}, \timeform{-28.90855D}) 
with an error of $\sim\timeform{0.3'}$.

For the PSF, we used a simulated radial profile of a point-like source.
We fitted the radial profile of G\,0.174$-$0.233 with 
the PSF plus a constant component model.
This point-source model is rejected with $\chi^{2}$/d.o.f = 36.4/10, 
as is shown in figure~\ref{fig:radpro-64cl}, 
and hence a single point-source origin is rejected. 
We already noted that the image shape is also unlikely to be a point source.
We therefore conclude that G\,0.174$-$0.233 is a diffuse source.

\subsubsection{G\,0.162$-$0.217}

G\,0.162$-$0.217 is located at $\sim\timeform{1.2'}$ west of G\,0.174$-$0.233.
Unlike G\,0.174$-$0.233, this source is not bright in the 6.4~keV line, 
but is visible in the continuum band (e.g. 4.0--6.0~keV). 
With the same method as is the case of HD\,316314, 
the peak position was determined to be 
$(\alpha, \delta)_{\rm J2000.0}$=(\timeform{266.71365D}, \timeform{-28.90989D}) 
with an error of $\sim\timeform{0.3'}$.

In figure~\ref{fig:image}c, we can see that the bright GCDX is extending near this source. 
Since the GCDX has strong K$\alpha$ lines of S\emissiontype{XV} (2.45~keV), 
Ar\emissiontype{XVII} (3.1~keV), Ca\emissiontype{XIX} (3.9~keV), and Fe\emissiontype{XXV} (6.70~keV), 
we eliminated the energy range below 4.0~keV and above 6.0~keV to suppress these emission lines. 

\begin{figure}[!ht]
  \begin{center}
  \FigureFile(80mm,80mm){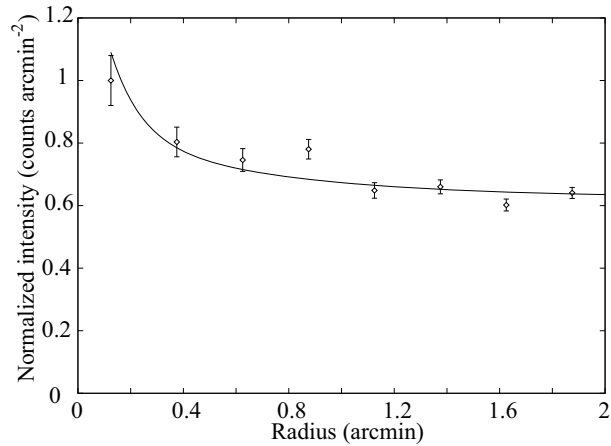}
  \end{center}
  \caption{Same as figure 2, but for G\,0.162$-$0.217 in the 4.0--6.0~keV energy band.
	Data inside \timeform{1'} from the peak position of G\,0.174$-$0.233 were excluded to reduce the contamination.
	The range of the profile was limited to inside \timeform{2'} to avoid a strong contribution from the GCDX.
  }
\label{fig:radpro-anocl}
\end{figure}

We checked any point source contribution to this source, 
while referring the Chandra catalog (Muno et al. 2006), and
found two point sources in the G\,0.162$-$0.217 region
(CXOGC\,J174651.9$-$285509 and CXOGC\,J174650.2-285451).
The total photon flux of these sources in the 0.5--8.0~keV energy band is 
5.4$\times10^{-6}$~cm$^{-2}$~s$^{-1}$, which is only $\sim 10\%$ of that 
of G\,0.162$-$0.217, and hence can be ignored.

Figure~\ref{fig:radpro-anocl} shows a radial profile of G\,0.162$-$0.217 in the 4.0--6.0~keV energy band,
where the flux in the \timeform{1'}-radius from the peak position of G\,0.174$-$0.233 was excluded.
To this profile, we fit the same model as G\,0.174$-$0.233, and found that
a point-source model is rejected with $\chi^2$/d.o.f = 16.9/6 (see figure~\ref{fig:radpro-anocl}).
The image shape (figure~\ref{fig:image}c) is also unlikely to be a point source.
One may argue that the radial profile of the source seems to be a point-source origin.
However, statistically the point source hypotheis is rejected 
at more than a 99\% confidence level ($\chi^2$/d.o.f = 16.9/6). 
Also, the morphology of G\,0.162$-$0.217 is asymmetrical and different 
from that of the point-like source of HD~316314 (figure~\ref{fig:image}a).
In fact, the point source flux reported by \citet{muno06} 
can only account for 10\% of the total flux of G\,0.162$-$0.217.
Accordingly, we regard that G\,0.162$-$0.217 is another extended source near the 6.4~keV clump G\,0.174$-$0.233.

\subsection{Spectra}

The X-ray spectra of the four sources (G\,0.174$-$0.233, G\,0.162$-$0.217, HD\,316314, and G\,0.30$-$0.20) 
were extracted from the source regions
(solid lines in figure~\ref{fig:image}), and we subtracted the background (dashed lines in figure~\ref{fig:image}).
The major background component is the GCDX, which includes a hot plasma ($k_{\rm B}T \sim 6.5$~keV) 
with a strong K$\alpha$ line of Fe\emissiontype{XXV} (6.7~keV).
We hence made the Fe\emissiontype{XXV}-K$\alpha$ line image (6.6--6.8~keV) and verified that 
no excess/deficient near at the source and background regions had been found. 
Since the flux of the GCDX depends on the distance from Sgr\,A$^*$ (\cite{koyama07b}; \cite{yamauchi90}), 
we selected the background regions near the sources, 
and nearly the same distance from Sgr\,A$^*$ as that of the source regions.

For the background-subtracted spectra, we merged the two FI (XIS\,0 and XIS\,3) data 
to increase the photon statistics,
because the redistribution matrix functions and the auxiliary response functions 
are essentially the same.
We generated redistribution matrix functions and the auxiliary response functions 
using \texttt{xisrmfgen} and \texttt{xissimarfgen} \citep{ishisaki07}. 
For the extended (diffuse) sources, we assumed a uniform flux distribution in the source region.
The model fittings were made for the FI and BI spectra simultaneously.
However, for simplicity in this paper, only the results of FI are shown for G\,0.174$-$0.233 and G\,0.162$-$0.217, 
because these two hard sources are more sensitive in the FI than the BI. 
Conversely, only the BI spectra are shown for the soft sources, HD\,316314 and G\,0.30$-$0.20.

\subsubsection{G\,0.174$-$0.233}

We extracted the XIS spectra of G\,0.174$-$0.233 from the source region of a $\sim\timeform{1.2'}$-radius solid 
circle shown in figure~\ref{fig:image}c. The background region is shown with the dashed annulus between radii of \timeform{2'} 
and \timeform{4'}, excluding HD\,316314 of the \timeform{1.5'} circle (figure~1c).
We fitted the background-subtracted spectra in the 2.0--10~keV energy band 
with an absorbed (\cite{morrison83}; \cite{anders89}) power-law plus two narrow Gaussian lines at 6.4~keV and 6.7~keV.
As a result, we found no significant 6.7~keV line with a flux upper limit 
of $\sim 3\%$ of that of the 6.4~keV line. 
We hence concluded that the GCDX background was properly subtracted.

\begin{figure}[!ht]
  \begin{center}
    \FigureFile(80mm,50mm){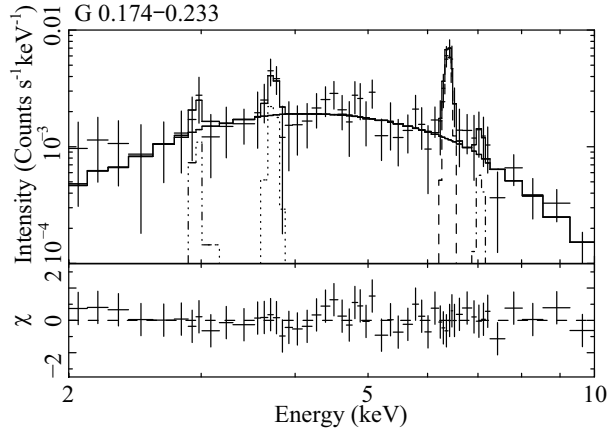}
  \end{center}
  \caption{Background-subtracted spectrum (FI) of G\,0.174$-$0.233.
    The data and the best-fit model are shown by the crosses and the solid line, respectively.
    The lower panels show the data residuals from the best-fit model.
	For brevity, only the FI spectrum is shown, although the fitting were simultaneous with the BI spectrum.
  }\label{fig:6.4keVspec}
\end{figure}

\begin{table*}[!ht]
  \begin{center}
   \caption{Best-fit spectral parameters of G\,0.174$-$0.233 and G\,0.162$-$0.217\footnotemark[$*$]}
   \label{tab:fitclumps}
    \begin{tabular}{lccc}
    \hline              
    Model component & Parameter (Unit) & G\,0.174$-$0.233 & G\,0.162$-$0.217 \\ 
    \hline
    Absorption & $N_{\rm H}$ ($10^{22}$~H~cm$^{-2}$) & 7.5$^{+2.0}_{-1.7}$ & 5.5$^{+0.8}_{-1.0}$ \\
    Power-law & $\Gamma$ & 1.7$^{+0.1}_{-0.2}$ & 2.5$^{+0.2}_{-0.6}$ \\
    Emission lines & & & \\
    \ \ Fe K$\alpha$ & Center (keV) & 6.40$^{+0.01}_{-0.02}$ & 6.39$^{+0.14}_{-0.12}$ \\
    & Flux\footnotemark[$\dagger$] & 6.2$^{+1.3}_{-1.2}$ & 1.9$^{+1.2}_{-1.0}$ \\
    & EW\footnotemark[$\ddagger$] & 0.95$^{+0.18}_{-0.19}$ & 0.36$^{+0.26}_{-0.19}$ \\
    \ \ Fe K$\beta$ & Center (keV) & 7.05\footnotemark[$\S$] & 7.04\footnotemark[$\S$] \\
    & Flux\footnotemark[$\dagger$] & 0.78\footnotemark[$\S$] & 0.23\footnotemark[$\S$] \\
    & EW\footnotemark[$\ddagger$] & 0.14$\pm 0.03$ & 0.06$\pm 0.04$ \\
    \ \ Ca K$\alpha$ & Center (keV) & 3.74$^{+0.03}_{-0.04}$ & --- \\
    & Flux\footnotemark[$\dagger$] & 2.9$^{+1.7}_{-1.5}$ & --- \\
    & EW\footnotemark[$\ddagger$] & 0.18$^{+0.10}_{-0.10}$ & --- \\
    \ \ Ar K$\alpha$ & Center (keV) & 2.96$^{+0.08}_{-0.25}$ & --- \\
    & Flux\footnotemark[$\dagger$] & $<$5.3 & --- \\
    & EW\footnotemark[$\ddagger$] & $<$0.22 & --- \\
    \hline
    Observed flux & $F_{\rm X}$\footnotemark[$\|$] & 4.8$\pm{0.4}$ & 4.2$\pm{0.3}$ \\
    Luminosity & $L_{\rm X}$\footnotemark[$\#$] & 3.7 & 3.2 \\
    \hline
    $\chi ^2$ / d.o.f & & 35.1 / 77 & 72.1 / 81 \\
    \hline
      \multicolumn{4}{@{}l@{}}{\hbox to 0pt{\parbox{120mm}{\footnotesize
	\par\noindent
	\footnotemark[$*$] The errors are at 90\% confidence level.\\
	\footnotemark[$\dagger$] Total flux ($10^{-6}$~photons~cm$^{-2}$~s$^{-1}$) in the line.\\
	\footnotemark[$\ddagger$] Equivalent width of the line in unit of~keV.\\
	\footnotemark[$\S$] The line energy and flux of Fe K$\beta$ are fixed at 
	110.3\% and 12.5\% \citep{kaastra93} of that of Fe K$\alpha$, respectively.\\
	\footnotemark[$\|$] The observed flux ($10^{-13}$~erg~cm$^{-2}$~s$^{-1}$) in the 2.0--10.0~keV energy band.\\
	\footnotemark[$\#$] The absorption-corrected X-ray luminosity ($10^{33}$~erg~s$^{-1}$) in the 2.0--10.0~keV energy band.
	 Distances to the sources are assumed to be 8~kpc.\\
	  }\hss}}	
    \end{tabular}
  \end{center}
\end{table*}

We then re-fit the spectra with a model of absorbed power-law plus two Gaussians 
for the K$\alpha$ and K$\beta$ lines of Fe\emissiontype{I} (6.40 and 7.06~keV).
However, we found two line-like residuals at $\sim3.7$~keV and $\sim3.0$~keV. 
These are close to the center energies of the K$\alpha$ lines of 
Ca\emissiontype{I} and Ar\emissiontype{I} (3.69 and 2.96~keV).
We therefore added two narrow Gaussians for these lines, and then
obtained a nice fit (figure~\ref{fig:6.4keVspec}).
In table~\ref{tab:fitclumps} we present the best-fit parameters.
Although the detection of the Ar-K$\alpha$ line is marginal (90\% confidence level),
the Ca-K$\alpha$ line was surely detected with $\sim 3\sigma$ significance.

\subsubsection{G\,0.162-0.217}

\begin{figure}[!ht]
  \begin{center}
    \FigureFile(80mm,50mm){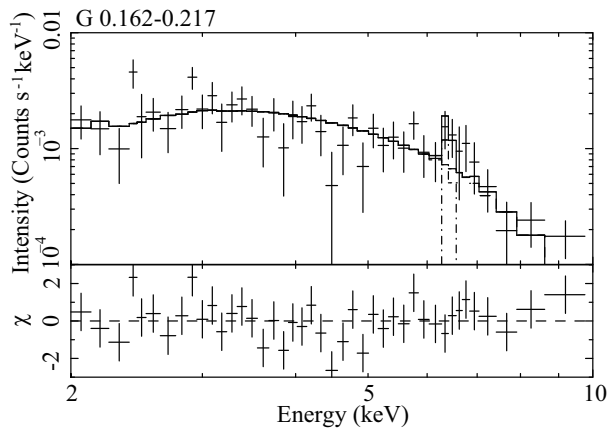}
  \end{center}
  \caption{Same as figure 4, but for the G\,0.162$-$0.217 spectrum.}
\label{fig:NWsp}
\end{figure}

We obtained the source spectra from the solid circle with a radius of $\sim\timeform{1.0'}$.
The background region and the spectral analysis procedures are the same as those of G\,0.174$-$0.233.
Since the background-subtracted spectra are similar to those of G\,0.174$-$0.233, 
we fitted with the same model, an absorbed power-law plus two narrow Gaussian line at 6.4~keV and 7.06~keV 
in the 2.0--10~keV energy band (figure~\ref{fig:NWsp}).
The best-fit parameters are given in table~\ref{tab:fitclumps}.
The spectra have weaker emission lines and steeper continuum than
those of G\,0.174$-$0.233, although the interstellar absorption is almost the same.

\subsubsection{The soft X-ray sources}

\begin{figure}[!ht]
  \begin{center}
    \FigureFile(80mm,50mm){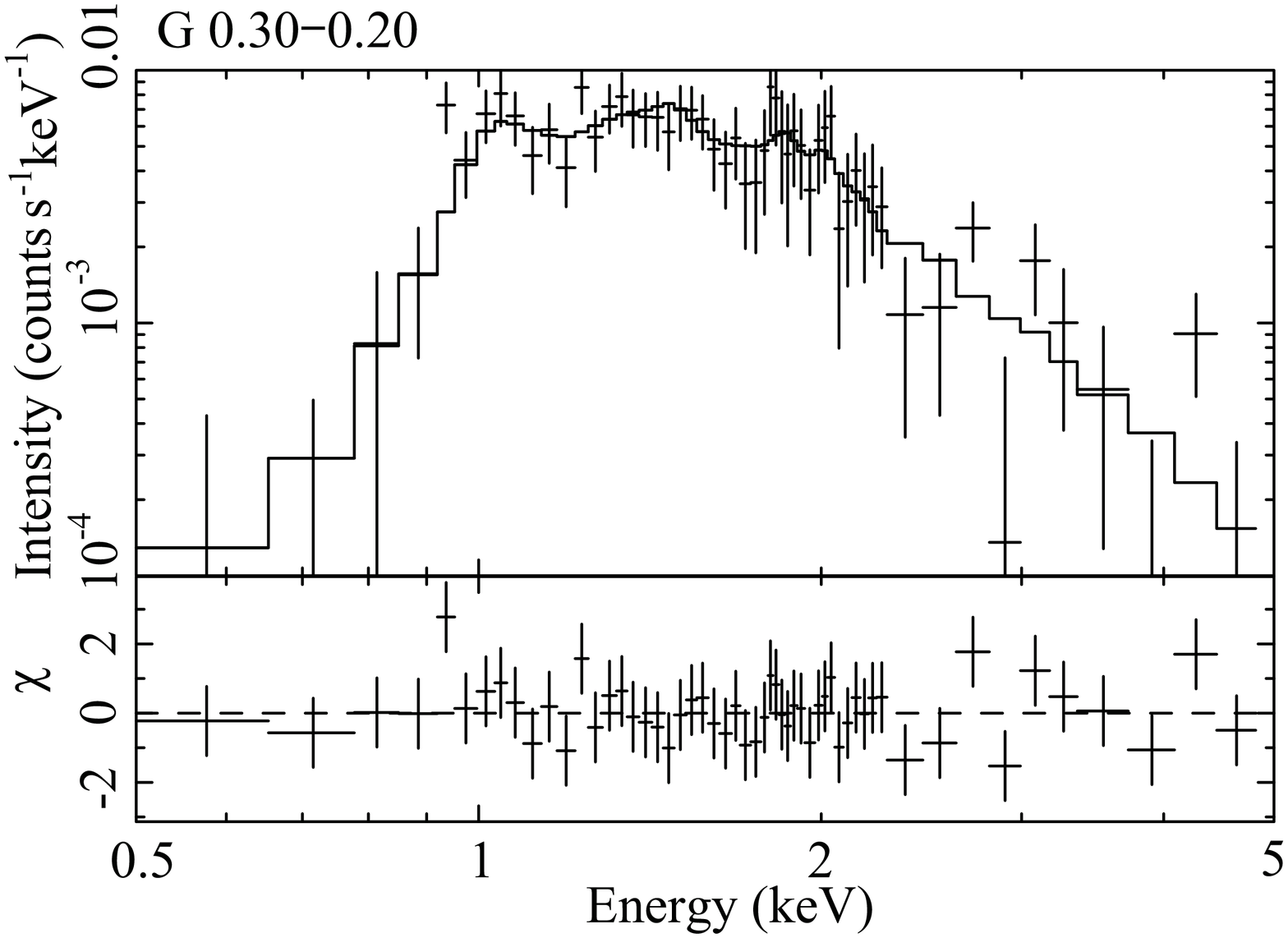}
  \end{center}
  \caption{Background-subtracted spectrum (BI) of G\,0.30$-$0.20.
	For brevity, only the BI spectrum is shown, although the fittings were simultaneous.
        The symbols are the same as those in figure~\ref{fig:6.4keVspec}.
  }\label{fig:clusterspec}
\end{figure}

\begin{table}[!ht]
  \begin{center}
   \caption{Best-fit spectral parameters of G\,0.30$-$0.20\footnotemark[$*$]}
   \label{tab:Cluster}
    \begin{tabular}{lcc}
    \hline              
    Model component & Parameter (Unit) & Value \\ 
    \hline
    Absorption & $N_{\rm H}$ ($10^{22}$~H~cm$^{-2}$) & 1.1$^{+0.16}_{-0.12}$  \\
    Thermal plasma & $k_{\rm B}T$ (keV) & 1.1$^{+0.14}_{-0.11}$ \\
    & $Z$ (solar\footnotemark[$\dagger$]) & 0.14$^{+0.05}_{-0.04}$ \\
    \hline
    Observed flux & $F_{\rm X}$\footnotemark[$\ddagger$] & 2.7$\pm 0.1$ \\
    \hline
    $\chi ^2$ / d.o.f. & & 74.8 / 99 \\
    \hline
      \multicolumn{3}{@{}l@{}}{\hbox to 0pt{\parbox{60mm}{\footnotesize
	\footnotemark[$*$] The errors are at 90\% confidence level.\\
	\footnotemark[$\dagger$] \citet{anders89}.\\
	\footnotemark[$\ddagger$] The observed flux ($10^{-13}$~erg~cm$^{-2}$~s$^{-1}$) in the 0.5--5.0keV energy band.\\
	  }\hss}}	
    \end{tabular}
  \end{center}
\end{table}

\begin{figure}[!ht]
  \begin{center}
    \FigureFile(80mm,50mm){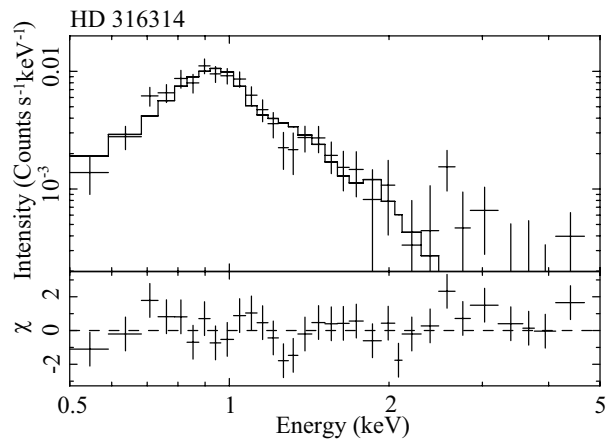}
  \end{center}
  \caption{Same as figure 6, but for the HD\,316314 spectrum.}
\label{fig:starspec}
\end{figure}

\begin{table*}[!ht]
  \begin{center}
   \caption{Best-fit spectral parameters of HD\,316314\footnotemark[$*$]}
   \label{tab:Star}
    \begin{tabular}{lcccc}
    \hline              
    Model component & Parameter (Unit) & Average & Quiescent state & Flare state \\ 
    \hline
    Absorption & $N_{\rm H}$ ($10^{21}$~H~cm$^{-2}$) & 1.3$^{+0.5}_{-0.4}$ & \multicolumn{2}{c}{1.3 (fixed)} \\
    Thermal plasma & $k_{\rm B}T$ (keV) & 0.80$\pm 0.04$ & 0.79$^{+0.05}_{-0.03}$ & 0.85$^{+0.11}_{-0.06}$ \\
    & $Z$ (solar\footnotemark[$\dagger$]) & 0.10$^{+0.03}_{-0.02}$ & \multicolumn{2}{c}{0.10 (fixed)} \\
    \hline
    Observed flux & $F_{\rm X}$\footnotemark[$\ddagger$] & 1.8$\pm 0.1$ & 1.6$^{+0.0}_{-0.1}$ & 3.0$^{+0.2}_{-0.3}$ \\
    \hline
    $\chi ^2$ / d.o.f. & & 75.9 / 64 & 71.8 / 66 & 36.6 / 43 \\
    \hline
      \multicolumn{5}{@{}l@{}}{\hbox to 0pt{\parbox{100mm}{\footnotesize
	\footnotemark[$*$] The errors are at 90\% confidence level.\\
	\footnotemark[$\dagger$] \citet{anders89}.\\
	\footnotemark[$\ddagger$] The observed flux ($10^{-13}$~erg~cm$^{-2}$~s$^{-1}$) in the 0.5--5.0keV energy band.\\
	  }\hss}}	
    \end{tabular}
  \end{center}
\end{table*}

\begin{figure}[!ht]
  \begin{center}
  \FigureFile(80mm,80mm){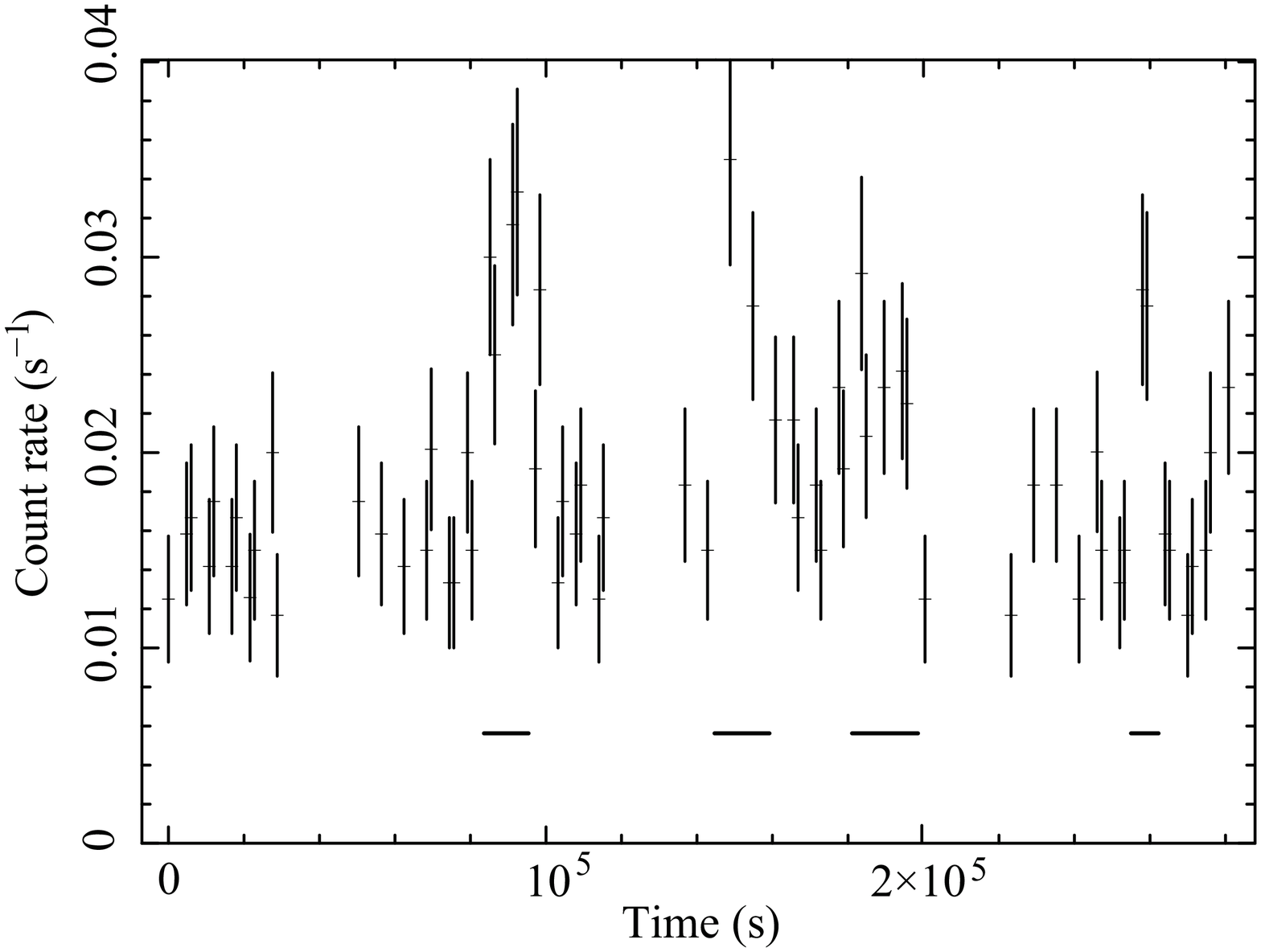}
  \end{center}
   \caption{XIS light-curve of HD\,316314 in the 0.5--1.5~keV energy band during the observation.
 	Three XIS data were co-added.
	We can see the X-ray activity of HD\,316314.
	We also show the time chosen to treat as the flare state.
	The under-line regions indicate the time of the flare state. 
	The other time regions are the quiescent state.
  }\label{fig:starlc}
\end{figure}

We extracted the spectra of G\,0.30$-$0.20 and HD\,316314 
from the solid ellipse and circle, respectively.
The background spectra were obtained from the regions 
between the inner solid (each source region) and outer dashed lines 
(figures~\ref{fig:image}a, and \ref{fig:image}b).
The background-subtracted spectra of G\,0.30$-$0.20 and HD\,316314 are shown 
in figure~\ref{fig:clusterspec} and figure~\ref{fig:starspec}, respectively.
The FI and BI spectra were simultaneously fitted in the 0.5--5.0~keV energy band 
with an absorbed thin-thermal plasma model (APEC in the XSPEC package; \cite{smith01}).
The best-fit model and parameters of G\,0.30$-$0.20 are shown 
in figure~\ref{fig:clusterspec} and table~\ref{tab:Cluster},
while those of HD\,316314 are shown in figure~\ref{fig:starspec} and table~\ref{tab:Star}.

We detected frequent flare-like events from HD\,316314, 
as shown in figure~\ref{fig:starlc}.
A constant flux hypothesis was rejected with $\chi^{2}$/d.o.f. = 111.9/65, 
while the background flux was fitted with a constant model 
($\chi^{2}$/d.o.f = 78.4/65).
We therefore regard that the flare-like activity is real.
By sorting the flux level (see under-line in figure 8), 
we constructed the spectra of the quiescent state (no under-line in figure 8) 
and the flare state (under-line in figure 8) and fitted with an APEC model.
We fixed the absorbing column and the metal abundance to the best-fit results of the average spectra.
The best-fit parameters are also listed in table~\ref{tab:Star}.
We find that the thermal plasma temperatures are almost the same between the two states, 
although the fluxes are different by a factor of two.

\section{Discussion}

\subsection{Nature of G\,0.174$-$0.233} 

The spectra of G\,0.174$-$0.233 have prominent emission lines representing  
K$\alpha$ of Fe and Ca. The best-fit center energy of the Fe-K$\alpha$ 
line corresponds to the theoretical value of a neutral iron.
Although the best-fit center energy of the Ca-K$\alpha$ line is slightly 
higher than the theoretical neutral calcium (table~\ref{tab:fitclumps}), 
the line is still consistent with that from the neutral
atoms, taking into account possible systematic gain uncertainty of $\sim10$~eV (see section 2).

The neutral lines indicate that the emission comes from a molecular cloud.
The absorbing column (table~\ref{tab:fitclumps}) is $\sim7.5\times10^{22}$~H~cm$^{-2}$, 
slightly larger than the typical value of $\sim6\times10^{22}$~H~cm$^{-2}$ in the 
GC region (\cite{reike89}; \cite{sakano02}). The absorption in the molecular
cloud is, therefore only a few times of 10$^{22}$~H~cm$^{-2}$. This means that the relevant 
molecular cloud, unlike Sgr\,B1, B2 and Sgr\,C, is a small-scale cloud.  

Two models for the diffuse X-ray emission from the molecular cloud have been proposed.
One is the irradiation of molecular clouds by external X-ray sources (the XRN model; \cite{koyama96}).
The other is the impact of LECRs on molecular clouds (the LECR model; \cite{yusef07}).

To discuss quantitatively, 
we compare the EW of the neutral K$\alpha$ line with the theoretical value.
With respect to the Fe\emissiontype{I}-K$\alpha$ line, the EW of the LECR model is 290~eV \citep{tatischeff03}.
They simulated the EW of the LECR model with the assumption of a solar abundance and 
the electrons population having a power-law distribution (photon index = 2) 
in the energy range of 10--100~keV.
The observed EW of $\sim950$~eV (table~\ref{tab:fitclumps}) for G\,0.174$-$0.233 
is too large for the LECR model by a factor of 3--4.
To fit the observation, a possible explanation for the LECR model is an over abundance by a factor of 3--4.
On the other hand, the XRN model is $\sim1$~keV for the solar abundance (e.g. \cite{murakami00}).
Thus, the XRN model does not require any over abundance, and hence an XRN scenario is more favored.

The Ca line is a remarkable feature of G\,0.174$-$0.233.
Such a strong Ca\emissiontype{I}-K$\alpha$ 
line has not been reported from any other XRNe.
The observed EW is 80--280~eV at the 90\% confidence level.
The LECR model expects an EW of $\sim10$~eV (derived from figure~7 in \cite{tatischeff03}),  
and is unlikely for the origin, since it needs a Ca abundance that has a factor of more than $\sim8$.
EW of the XRN model is estimated to be 50--60~eV.
Although we need the Ca abundance to be a factor more than $\sim1.5$, 
the XRN model is more likely for the origin than the LECR model.

A past X-ray flare of Sgr\,A$^*$ was proposed as an irradiating source of many other XRNe 
(e.g. \cite{koyama96}). To apply this scenario to G\,0.174$-$0.233,
the past (300~years before) outburst 
of 1--2$\times10^{39}$~erg~s$^{-1}$, should have been lasting until 150~years ago, 
because the projected distance of G\,0.174$-$0.233 from Sgr\,A$^*$ 
is about half (20') of the other XRNe (e.g. \cite{nobukawa08}).
This requirement is similar to 
that for the other XRN near the Radio Arc (\cite{koyama09}). 

A low-mass X-ray binary, SAX\,J1747.0$-$2853, exhibited yearly outbursts between 1998--2001. 
The burst peak was $\sim3\times10^{37}$~erg~s$^{-1}$ (at 8~kpc) in 2000 \citep{werner04}. 
Since this transient source is close to G\,0.174$-$0.233 (separated by only \timeform{2'}), 
one may argue that it is a potential candidate for the illuminating source of G\,0.174$-$0.233.
Taking account of a positional error of $\sim\timeform{0.3'}$, 
the lower limit on the separation of G\,0.174$-$0.233 and SAX\,J1747.0$-$2853 
is given by a projected distance, and is estimated to be $\sim4$~pc of the light travel time of ~12 years. 
A putative flare responsible for the Suzaku 6.4 keV emission can be constrained
by X-ray observations in 1998-2001 with BeppoSAX, ASCA, Chandra, and XMM-Newton, 
because the separation between these observations and Suzaku is roughly 10~years.
However, SAX\,J1747.0$-$2853 should be as bright as $\sim3\times10^{37}$~erg~s$^{-1}$ for more than a few years. 
The averaged luminosity over a few years  was smaller than $\sim10^{36}$~erg~s$^{-1}$, too faint to be a candidate.
In fact, \citet{werner04} reported that the durations of the outbursts were between one week and a few months.
Since the size of G\,0.174$-$0.233 is at least a few light-years,  
SAX J1747.0$-$2853 could be an irradiating source only if the fluorescing gas (G 0.174$-$0.233) 
were to be concentrated within a sheet of, at most, 
a few light-months thickness perpendicular to the direction of SAXJ1747.0$-$2853.
This is unlikely, although we can not rule out the possibility.

\subsection{Nature of G\,0.162$-$0.217 }

The absorbing column of G\,0.162$-$0.217 (table~\ref{tab:fitclumps}) 
is consistent with the typical value of $\sim6\times10^{22}$~H~cm$^{-2}$ in the 
GC region (\cite{reike89}; \cite{sakano02}).
Therefore, this source is likely to be near the GC.
Since G\,0.174$-$0.233 is also very likely to be located at the GC, the real separation
of these two sources would be very small; the least case is $\sim\timeform{1.2'}$ or 3--4~pc.
Since this source is only $\sim\timeform{1.2'}$ from the 
strong 6.4~keV-line source G\,0.174$-$0.233,
we estimated possible 6.4~keV-line contamination from G\,0.174$-$0.233, assuming
that the flux distribution is in between uniform in all of the source region, 
and is concentrated at its center.
As a result, the range of possible contamination was found to be 15--29\% of G\,0.174$-$0.233.
Then, the 6.4~keV flux contamination is estimated to be $\sim20$\%, 
and hence the contamination-removed EW of G\,0.162$-$0.217 is $\sim0.2$~keV.
We thus regard that the neutral iron K-shell line is present in G\,0.162$-$0.217.

\begin{figure}[!ht]
  \begin{center}
    \FigureFile(80mm,80mm){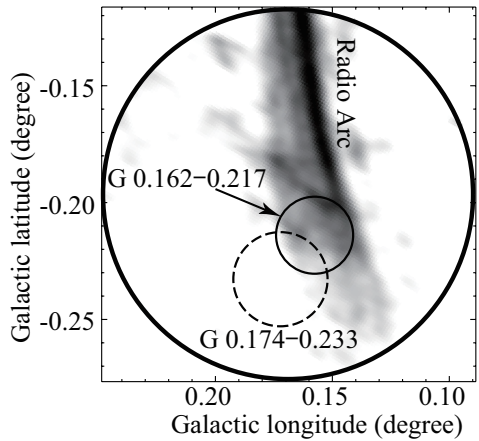}
  \end{center}
  \caption{A 4.735~GHz radio continuum map 
from the National Radio Astronomy Observatory / Very Large Array archive survey image
. 
The Very Large Array field is shown with the thick solid circle.
The positions of G\,0.174$-$0.233 and G\,0.162$-$0.217 are shown by
the dashed and thin-solid circles on the Radio Arc.
The coordinates are the Galactic longitude and latitude.}
\label{fig:Radio-Arc-foot}
\end{figure}

The X-ray origin is uncertain but, interestingly, G\,0.162$-$0.217 
is located at adjacent to the south end of the Radio Arc (\cite{larosa00}; \cite{yusef04}), 
as shown in figure~\ref{fig:Radio-Arc-foot}.
\footnote{The National Radio Astronomy Observatory is a facility of the National Science Foundation operated 
under cooperative agreement by Associated Universities, Inc.}. 
Since the Radio Arc is a site of relativistic electrons, which may also include LECR,
the electrons would come along the magnetic field line of the Radio Arc. 
Thus, it may be conceivable that the X-rays of G\,0.162$-$0.217 are due to the LECRs.
The observed EW of Fe\emissiontype{I} K$\alpha$ of $\sim0.2$~keV is consistent 
with the LECR model (EW$\sim0.3$~keV).

\subsection{The Soft Sources}

The source region of G\,0.30$-$0.20 includes two stellar clusters, 
DB\,00-5 and DB\,00-6, in X-rays, which are the members of the H\emissiontype{II} 
region Sh2-20 (\cite{dutra03}, \cite{law04}).
\citet{law04} found several X-ray point sources in these two stellar clusters
(crosses in figure~\ref{fig:image}b), with
a total flux of $3.2\times10^{-5}$~counts~cm$^{-2}$~s$^{-1}$. 
On the other hand, the Suzaku flux was $\sim9\times10^{-5}$~counts~cm$^{-2}$~s$^{-1}$, 
which is three times larger than that of Chandra.
Chandra is sensitive for point sources, due to a superior spatial resolution, 
and hence it can resolve point sources, and estimate the flux due to them.
The rest of the flux would have a diffuse origin, such as emission from the H\emissiontype{II} region.
This argument is supported by NTT observations of DB\,00-5 and DB\,00-6 (see 
Figure~2 of \cite{dutra03}), which show diffuse features around each stellar cluster.
Chandra found no extended emission from this region, simply because 
a largely extended source is difficult to detect with Chandra. 
The plasma temperature is k$T\sim1$~keV.
This is significantly higher than the typical temperature of diffuse X-rays in the other H\emissiontype{II} 
regions, although only a limited
sample of the diffuse X-rays from stellar cluster is presently available.   

The Chandra X-ray source in DB\,00-6 was reported to have an absorbing column of 
0.6--1.6$\times10^{22}$~H~cm$^{-1}$, which is consistent with infrared observations, and
is approximately the same value as that of DB\,00-5. 
Suzaku determined a more accurate absorbing column for the entire region of
G\,0.30$-$0.20 to be 1.0--1.3 $\times10^{22}$~H~cm$^{-1}$.  Assuming the average
density toward this source to be 1~H~cm$^{-3}$ (e.g. \cite{jenkins76}), 
the distance is estimated to be 3~kpc.
The X-ray luminosity is $\sim1\times10^{33}$~erg~s$^{-1}$, 
which may be classified an X-ray faint star cluster. 

The soft X-ray point source HD\,316314 has been identified as being a F0 star.
It exhibited a long-term time variability in Chandra observations. 
However, the averaged fluxes for Chandra and Suzaku 
are almost the same as $\sim1.1\times10^{-4}$~counts~cm$^{-2}$~s$^{-1}$. 
With the same assumption of G\,0.30$-$0.20, the distance of HD\,316314 is
estimated to be 300~pc, and thus the luminosity is  $\sim2\times10^{30}$~erg~s$^{-1}$.
This luminosity is very high as a low-mass star of F0, but is not exceptional if
this star is young. Since low-mass stars exhibit stellar flares 
due to magnetic activity, the flare-like events from this source may be natural.
The flare spectrum, however, showed no significant hardening, which is
unnatural for magnetic activity. The origin of the X-rays is therefore 
still debatable.
\\
\par
We are grateful to all members of the Suzaku hardware and software teams and the science 
working group, especially H.~Matsumoto and M.~Nobukawa, for their useful comments and supports.
We also thank Y.~Hyodo, H.~Uchiyama, and M.~Sawada for improving the draft quality.
This work was supported by a Grant-in-Aid for the Global COE Program
 ``The Next Generation of Physics, Spun from Universality and Emergence'' from the Ministry 
of Education, Culture, Sports, Science and Technology (MEXT) of Japan. 
This work was also supported by Grants-in-Aid of Ministry of Education, Culture, Sports, 
Science and Technology (No.\,18204015 and 20340043).

\end{document}